\documentclass[11pt]{article}

\usepackage[iso-8859-7]{inputenc}
\usepackage{epsfig}
\usepackage{graphicx}
\usepackage{epstopdf}
\usepackage{amsfonts}
\usepackage{amssymb}
\usepackage{amsthm}
\usepackage{amsmath}
\usepackage{multirow}

\newcommand{\newc}{\newcommand}
\newc{\ra}{\rightarrow}
\newc{\lra}{\leftrightarrow}
\newc{\be}{\begin{equation}}
\newc{\ee}{\end{equation}}
\newc{\bs}{\begin{split}}
	\newc{\es}{\end{split}}
\newc{\ba}{\begin{eqnarray}}
\newc{\ea}{\end{eqnarray}}
\newc{\ov}{\overline}
\newc{\pa}{\partial}
\newc{\D}{\Delta}

\newc{\nn}{\nonumber}

\begin{document}
	\begin{titlepage}
		
		%\vspace*{-15mm}
		%\begin{flushright}
		%SHEP-11-XX\\
		%\end{flushright}
		\vspace*{0.7cm}

		\begin{center}
			{\Large {\bf Light Sterile Neutrino and Low Scale Left-Right Symmetry in D-brane inspired $SU(4)_C \times SU(2)_L \times SU(2)_R$}}
			\\[12mm]
			George K. Leontaris$^{a}$
			\footnote{E-mail: \texttt{leonta@uoi.gr}} and
			Qaisar Shafi$^{b}$
			\footnote{E-mail: \texttt{shafi@bartol.udel.edu}}
			\\[-2mm]
			
		\end{center}
		\vspace*{0.50cm}
		\centerline{$^{a}$ \it
			Physics Department, Theory Division, Ioannina University,}
		\centerline{\it
			GR-45110 Ioannina, Greece }
		\vspace*{0.2cm}
		\centerline{$^{b}$ \it
			Bartol Research Institute, Department of Physics and Astronomy, University of Delaware,}
		\centerline{\it
			DE 19716,  Newark, USA}
		\vspace*{1.20cm}
		
		\begin{abstract}
			\noindent
			
			Motivated by the ongoing searches for new physics at the LHC, we explore the low energy consequences of a D-brane inspired
			$ SU(4)_C\times  SU(2)_L \times  SU(2)_R$ (4-2-2) model. The Higgs	sector consists of an $SU(4)$ adjoint, a pair $H+\bar H$
			 in $(4,1,2)+(\bar 4,1,2)$, and a bidoublet field in $h(1,2,2)$. With the $SU(4)$ adjoint the symmetry breaks to a left-right 
			 symmetric $SU(3)_C\times U(1)_{B-L} \times SU(2)_L \times SU(2)_R$ model. A missing partner mechanism protects the $SU(2)_R$
			 Higgs doublets in $H,\bar H$, which subsequently break the  symmetry to the Standard Model at a few TeV scale. An inverse 
			 seesaw mechanism generates masses for the observed neutrinos and also yields a sterile neutrino which can play the r\^ole of 
			 dark matter if its mass lies in the keV range. Other phenomenological implications including proton decay are  briefly discussed.

		\end{abstract}
		
\end{titlepage}

	% % % % % % % % % % % % % % % % % % % %

\section{Introduction}

Spontaneously broken left-right symmetry and the idea of quark-lepton unification are nicely encapsulated in the $SU(4)_C\times SU(2)_L\times SU(2)_R$ (4-2-2) gauge symmetry proposed by Pati and Salam~\cite{Pati:1974yy}. With gauge bosons that can mediate proton decay absent, the 4-2-2 symmetry may
 well be broken at scales that are orders of magnitude lower than the usual grand unified scale $M_{GUT} \simeq 2 \times  10^{16}$ GeV. 
A lower bound on the 4-2-2 breaking scale of around 100 TeV  arises in the simplest model from the non-observation of the decay 
$K \to \mu + e $ mediated by some of the gauge bosons in $SU(4)_C$. The experimental constraints on the left-right gauge symmetry  breaking 
scale are significantly milder~\cite{Sirunyan:2017acf,ATLAS:2015nsi}, and indeed this symmetry may be broken at a scale that may be accessible
 either at the LHC or its upgrades. 

In this letter we propose to explore the low energy consequences of a 4-2-2 model derived from an intersecting D-brane framework.
As is well known, the 4-2-2 model realises quark-lepton unification in a natural way.  Several other  attributes of this symmetry~\cite{Mohapatra:1974hk}-\cite{Lazarides:1980nt} are the incorporation of the right-handed  neutrino in the spectrum, 
the absence of gauge bosons mediating fast proton decay etc.\footnote{For a recent review see~\cite{Pati:2017ysg}.} The last 
few years there has been growing  theoretical and experimental interest  in the theory of fundamental interactions in the TeV region.
We believe that the 4-2-2 model incorporates all the ingredients to interpret possible related findings and therefore
  in this note, we would like to 
 examine low scale  symmetry breaking patterns  and explore possible predictions.
Indeed, the present experimental bounds on scalar superpartners are close to the TeV scale while searches  for new gauge bosons have put
lower limits at a few TeV.  Supersymmetric scalar masses of the MSSM states are definitely related to supersymmetry  breaking,
 while the existence of new gauge bosons are naturally associated with some new symmetry breaking scale.
A natural candidate of such a new gauge symmetry, not far above the electroweak (EW) scale, is $SU(2)_R$, so
that above this breaking scale, the model is left-right symmetric.  Moreover, there is a good chance  that the
associated gauge boson with a  relatively low $SU(2)_R$ breaking scale   leaves its signature in experiments
through new interactions, which may be identified in future Diboson searches~\cite{CMS:2017wsr}.
The left-right symmetric model is naturally embedded in a quark-lepton unified (4-2-2) symmetry which admits an
interesting  string realisation in the context of heterotic superstrings, the intersecting 
D-brane scenarios, as well as in F-theory~\cite{Antoniadis:1988cm}-\cite{Cvetic:2015txa}. However, although the PS symmetry 
unifies quarks and leptons, yet its symmetry structure consists of a product of 
non-abelian factors and as such, does  not necessarily imply unification of the gauge couplings. Given this fact, 
from the above string theory realisations, 
the intersecting  D-brane set up, appears to be the most natural scenario.   Indeed, within this context, each group 
factor is associated with a different brane stack 
with its own gauge coupling strength, therefore the PS model is naturally  motivated in an intersecting D-brane framework. 
Moreover, the  tight connection between the unification scale and the Planck scale $M_{Planck}$ is no longer 
mandatory and the symmetry breaking scale can be substantially lower than $M_{Planck}$. 

In this note, we investigate the realisation of a two step symmetry breaking scenario of the PS model
built  in the context of open string theories and D-branes.  At the first stage the PS symmetry breaks 
at $M_{GUT}$ to the left-right (LR) symmetric model with the use of the adjoint Higgs. At a second scale $M_R\ll M_{GUT}$
 a pair of righ-handed Higgs doublets triggers the breaking of the LR-symmetry down to the Standard Model one.

 The layout of the paper is as follows. In section 2 we discuss the D-brane framework for the 4-2-2 model and present the particle spectrum
  in Table 1. A variety of new fields appear that are absent in the standard field theory constructions. They include a SM sterile neutrino
   which turns out to be a plausible dark matter candidate. In section 3 we consider the two step breaking of 4-2-2 to the SM.  We exploit here a {\it triplet-singlet} splitting mechanism, which is analogous to the well-known {\it doublet-triplet} mechanism of $SU(5)$. The left-right symmetry breaking scale 
   lies in the TeV range which may be found at the LHC or its future upgrades. Section 4 contains a discussion of neutrino masses which includes 
   the inverse seesaw mechanism and a sterile keV mass neutrino. A brief discussion in this section of proton decay shows that it is adequately 
   suppressed in this class of models. Our summary and conclusions are presented in section 5.

\section{4-2-2 Spectrum from intersecting D-branes}

In this section we will present the basic features of an open string realisation of the 4-2-2 gauge symmetry.
In general, there are several methods to construct 4-2-2 vacua in string theory, such as Calabi-Yau
compactification, Type IIA string theory, interacting CFT constructions and also Gepner constructions.
For the purposes of this work we find it  convenient to represent these models in terms of  intersecting  D-brane
configurations which is the appropriate description for type IIA string theory.

\begin{figure}[!bth]
  \centering
  \includegraphics[scale=1.5]{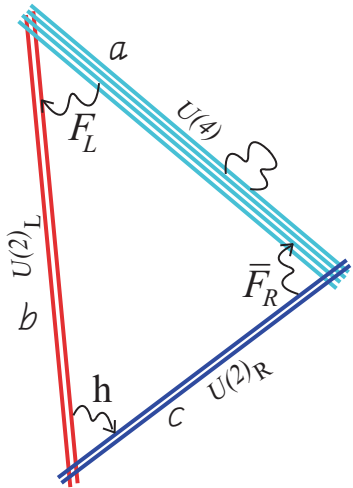}
  \caption{Intersecting D-brane configuration for the 4-2-2 symmetry. The letters $a,b,c$ respectively
  stand for the $U(4)_C, U(2)_L$ and $U(2)_R$ brane stacks. Under orientifold planes
 there are the mirror branes (not shown in this figure) denoted with $a^*, b^*, c^*$ in the main text.}
  \label{f1}
  \end{figure}

The low energy phenomenology of the 4-2-2 symmetry built in the framework of intersecting D-brane scenaria
 differs in many respects from the corresponding model which admits an $SO(10)$ embedding~\cite{Lazarides:1980nt}.
The main reason is that intersecting  D-brane  constructions yield  an $U(4)\times U(2)\times U(2)$
gauge group, which is  the 4-2-2 symmetry augmented by three $U(1)$ factors. Indeed,
recalling  that  $U(n)\simeq SU(n)\times U(1)/Z_n$ the final gauge symmetry of the D-brane version of
the effective field theory model is
\ba
  SU(4)_C\times SU(2)_L \times SU(2)_R\times U(1)_C\times U(1)_L\times U(1)_R\,\cdot\label{PSGS}
\ea
The intersecting D-brane set up associated with the above symmetry is depicted in Figure~\ref{f1}. Each gauge group  factor
$U(n)$ is associated with a set of $n$ parallel, almost coincident, D-branes, while the various massless states are represented
by open strings attached on the various sets of branes in the appropriate configurations. This set up gives rise to   bi-fundamental
representations which accommodate the particle spectrum of the model.  The most general picture of the available representations
 is as follows:  i) Open string connecting two brane stacks, give rise to bifundamental representations
with respect to the corresponding gauge groups;  ii) open strings stretched between a D-brane and its image transform in the antisymmetric
or symmetric representation of the gauge group; ii)  for each group factor, we should include the adjoint representation
which arises from open strings stretched between branes in the same stack.

To underline the salient features of the D-brane derived 4-2-2 models,
in Table~\ref{T1} we present a minimal set of fields obtained in the context
of a D-brane set up, which are required for the realisation of a viable effective model. There
are states originating from string vibrating at the intersections $ab, bc, ca$
of the three D-brane stacks $a,b,c$. There are also states emerging from strings
with ends attached to the mirror branes $a^*, b^*, c^*$ (not shown in figure).
 The  observed chirality of the spectrum can be adjusted from appropriate string
boundary conditions but, in general, the spectrum may contain additional
vector like pairs. 

As expected, all massless states associated with the open string spectrum, are
charged under the  extra $U(1)$ factors.  Indeed, returning to the content of Table~\ref{T1},
 we note that
the last three entries show the `charges' of the representations  under the
three abelian factors $U(1)_C, U(1)_L, U(1)_R$.
 It is  remarkable  that the abelian factor $ U(1)_C$  plays a central r\^ole in defining the baryon and
lepton  quantum numbers.  Indeed, if we consider the breaking
\be
U(4)_C \to SU(4)_C\times U(1)_C \to SU(3)\times U(1)_{B-L}\times U(1)_C\,,
\ee
we find out that they are defined in terms of the linear
 combinations~\cite{Leontaris:2000hh}
 \ba
   U(1)_{\cal B} &=&  \frac 14 \left( U(1)_C+U(1)_{B-L}\right)\label{Bdef}\,,\\
   U(1)_{\cal L} &=&  \frac 14 \left( U(1)_C-3U(1)_{B-L}\right)\,\cdot\label{Ldef}
   \ea
  
  \noindent  
 We note that in the context of the Standrad Model these are global symmetries
which  are anomalous. In the framework of intersecting D-brane constructions,
the $U(1)_{C,L,R}$ symmetries also are anomalous and only particular combinations
are anomaly-free. It transpires that these $U(1)$ anomalies  are cancelled by the generalised
Green-Schwarz mechanism (through couplings between  the RR two form field and the
corresponding field strength, see for example~\cite{Ibanez:2001nd}),
 which induces masses for the corresponding gauge bosons.
Yet, at the perturbative level, these remain as global symmetries  which prevent
rapid baryon and lepton number violating processes.

In the following we describe the embedding of the Standard Model states in the 4-2-2 representations
 and the Higgs mechanism for a two-stage symmetry breaking to SM.
The fermion generations  are accommodated in the following representations
  \ba
F_L+\bar F_R=(4,2,1)+(\bar 4,1,2)\,,
  \ea
which make up just the   $16$ of the  $SO(10)$.
\begin{table}[!b]
\centering
\renewcommand{\arraystretch}{1.2}
\begin{tabular}{llccc}
\hline
Intersection & $SU(4)_C\times SU(2)_L\times SU(2)_R$& $Q_C$ & $Q_{2L}$ & $Q_{2R}$ \\
\hline
 $ab$   & $3\times F_L\,(4,\bar 2,1)$ & $1$ & $-1$ & $0$    \\
   \hline
$ac$& $3\times \bar F_R\,(\bar 4,1,2)$ & $-1$ & $0$  & $1$ \\
\hline
$ac^*$& $\bar H\,(\bar 4,1,\bar 2)$ & $-1$ & $0$  & $-1$ \\
& $ H\,(4,1, 2)$ & $1$ & $0$  & $1$ \\
\hline
 $aa^*$   & $ S^{\pm}_{10}\,(10,1,1)$ & $ \pm 2$ & $0$ & $0$        \\
  & $ D^{\pm}_6\,(6,1,1)$ & $\pm 2$ & $0$ & $0$        \\
 $cc^*$   & $ \Delta_R\,(1,1,3)$ & $0$ & $0$ & $\pm 2$        \\
  & $  (N,\bar N)$-singlets&0&0&$\pm 2$\\
 $bb^*$  & $ \Delta_L\,(1,3,1)$ & $0$ & $\pm 2$  & $0$       \\
   & $(\nu_s,\bar \nu_s)$-singlets & $0$ & $\pm 2$ & $0$        \\
\hline
 $bc$ & $h(1,2,\bar 2)$ & $0$ & $1$ & $-1$        \\
  & $\bar h(1,\bar  2,2)$ & $0$ & $-1$ & $1$        \\
   $bc^*$   & $\left.\begin{array}{c}h'(1,2,2)\\
                   \bar{h}'(1,\bar 2,\bar 2)
                   \end{array}\right.$  & $\left.\begin{array}{c}0\\
                   0
                   \end{array}\right.$ &$\left.\begin{array}{c}1\\
                   -1
                   \end{array}\right.$& $\left.\begin{array}{c}1\\
                   -1
                   \end{array}\right.$       \\
\hline
 $ aa$   & $ \Sigma\,(15, 1,1)$ & $0$ & $0$ & $0$    \\
  $ bb$   & $ \Sigma_{3_L}\,(1, 3,1)$ & $0$ & $0$ & $0$    \\
   $ cc$   & $ \Sigma_{3_R}\,(1, 1,3)$ & $0$ & $0$ & $0$    \\
\end{tabular}
 \caption{Minimal Spectrum and the corresponding quantum numbers that emerge
 in a D-brane configuration with $U(4)_C\times U(2)_L\times U(2)_R$ gauge symmetry.
 \label{T1} }
\end{table}
The Standard Model particle assignment is
  \ba
F_L&=&({4},{2},{1})
=Q({3},{2},\frac 16) + \ell{({1},{2},-\frac 12)}\,,\nonumber\\
\bar{F}_R&=&(\bar{{4}},{1},{2})
=u^c({\bar 3},
{1},-\frac 23)+d^c({\bar3},{1},\frac 13)+
                           \nu^c({1},{1},0)+ e^c({1},
                           {1},1)\,\cdot\label{fermions}
\ea
    The Higgs sector comprises the following fields.
    The non-trivial representations are the $SU(4)_C$ adjoint
    Higgs field $\Sigma=(15,1,1)$, the bidoublets $h,h'$ in $(1,2,2)+c.c.$,
     and the two Higgs fields $\bar H$ and $ H$, namely
 \ba
 \bar H=(\bar 4,1,2)&=&(\bar Q_H+\bar L_H)\to (u^c_H,\,d^c_H,\,e^c_H,\,\nu^c_H)\label{H4b}
 \\
 H=( 4,1,2)&=&( Q_H+ L_H)\to (\bar u^c_H,\,\bar d^c_H,\,\bar e^c_H,\,\bar \nu^c_H)~\cdot\label{H4}
 \ea
In an $SO(10)$ embedding we note that $\bar H, H$ descend from the $16$ and $\overline{16}$ of $SO(10)$ respectively.
In addition, a suitable set of singlet fields  develop vacuum expectation values (vevs)
 to provide  masses for the vector-like pairs and right-handed neutrinos.
The minimal spectrum derived from the D-brane scenario is given in Table~\ref{T1}, where
the transformation properties under the various quantum numbers are also shown. Notice, in
particular, that the triplet pair $D_3^-+\bar D_3^+$ found in the decomposition
of the sextets has the quantum numbers of leptoquarks since they 
simultaneously carry both  baryon and lepton
 number under the definitions~(\ref{Bdef},\ref{Ldef}).

\section{ Two Step Spontaneous Symmetry Breaking}

Having defined the spectrum of the model,  in this section we proceed with  the
implementation of the two-step spontaneous breaking of 4-2-2 symmetry.
 The $SU(4)_C\to  SU(3)_C\times U(1)_{B-L}$ breaking is realised with the Higgs field
 $\Sigma=(15,1,1)$  listed in Table 1:
 \ba
 (15,1,1)\ra \langle\Sigma\rangle &=&{\tiny \left(
 \begin{array}{cccc}
 \frac V3 &0 & 0 &0\\
 0 &\frac V3 & 0&0 \\
 0 & 0 & \frac V3&0\\
 0 & 0 &0& -V
 \end{array}
 \right)} \cdot\label{Ajvev}
 \ea
Then, at this stage, the  symmetry of the model is
\be
 SU(3)_C\times SU(2)_L\times SU(2)_R\times  U(1)_{B-L}\times U(1)_L\times U(1)_R\,\cdot\label{LRSYM}
 \ee
The decomposition of the various representations of the model are shown in Table~\ref{T2}.

The breaking of the left-right symmetric group in (\ref{LRSYM}) to the SM gauge group  takes place
from the non-zero vevs along the neutral components  $\langle\nu_H^c\rangle,\langle\bar\nu_H^c\rangle$
of  the right-handed doublet fields $L_H, \bar L_H$ in $H,\bar H$.  Notice that these Higgs fields  
 are also `charged' under $U(1)_R$. Then, their vevs, combined with the $\bar N, \nu_s$ singlet non-zero vevs 
 which will be discussed later, will also break the $U(1)_L$ and $ U(1)_R$ symmetries. At the final stage, the  SM symmetry
 breaking occurs with a non-zero vev  of the bidoublet $h$.

\begin{table}[!b]
\centering
\renewcommand{\arraystretch}{1.2}
\begin{tabular}{c|ccccc}
\hline
 $SU(3)_C\times SU(2)_L\times SU(2)_R$& $Q_{B-L}$ & $Q_{2L}$ & $Q_{2R}$ &${\cal B}$&${\cal L}$\\
\hline
  $3\times Q_L\,(4, 2,1)$     & $\frac 13$ & $-1$ & $0$ &$\frac 13$&$0$   \\
        $3\times L\,( 1, 2,1)$      & $-1$       & $-1$ & $0$ &$0$&$1$    \\
   \hline
 $3\times  Q^c_R\,(\bar 3,1,2)$ & $-\frac 13$ & $0$  & $1$&$-\frac 13$&$0$  \\
 $3\times  L^c_R\,( 1,1,2)$     & $1$ & $0$  & $1$&$0$&$-1$  \\
\hline
  $ Q_H\,(3,1, 2)$        & $\frac 13$  & $0$  & $1$&$\frac 13$&$0$ \\
 $ L_H\,( 1,1, 2)$ & $-1$ & $0$  & $-1$&$0$&$1$  \\
 $\bar Q_H\,(\bar 3,1, 2)$ & $-\frac 13$ & $0$  & $-1$&$-\frac 13$&$0$  \\
 $\bar L_H\,( 1,1, 2)$ & $1$ & $0$  & $1$&$0$&$-1$  \\
 \hline
 $ D_3^{+}\,(3,1,1)$ & $\frac 23$ & $0$ & $0$   &$\frac 23$&$0$      \\
  $\bar D_3^{+}\,(\bar 3,1,1)$ & $ \frac 23$ & $0$ & $0$   &$\frac 13$&$1$      \\
   $ D_3^{-}\,(3,1,1)$ & $-\frac 23$ & $0$ & $0$    &$-\frac 13$&$-1$     \\
  $ \bar D_3^{-}\,(\bar 3,1,1)$ & $- \frac 23$ & $0$ & $0$   &$-\frac 23$&$0$        \\
 $ \Delta_R\,(1,1,3)$ & $0$ & $0$ & $\pm 2$       &$-$&$-$  \\
 $  (N,\bar N)$&0&0&$\pm 2$&$-$&$-$ \\
 $ \Delta_L\,(1,3,1)$ & $0$ & $\pm 2$  & $0$    &$-$&$-$    \\
$(\nu_s,\bar\nu_s)$ & $0$ & $\pm 2$ & $0$    &$-$&$-$
 \\
    $h,h'(1,2,2)$ & $0$ & $\pm 1$ & $\mp 1$   &$-$&$-$    \\
   \hline
\end{tabular}
 \caption{Spectrum of the left-right symmetric model  after the breaking of the 4-2-2 symmetry.
 \label{T2} }
\end{table}

Before studying some implications for the effective low energy theory, recall first that in the D-brane construction 
 described above, in addition to the chiral states $F_L, \bar F_R$ accommodating the fermion generations,
vector-like pairs such as $F^i_L+\bar F^i_L$,  $F^i_R+\bar F^i_R$,  $D^i_6+\bar D^i_6$, $i=1,2,\dots$ etc.,
are usually present
in the zero mode spectrum. However, such pairs arise with opposite $U(1)$ charges and, in principle,
receive heavy masses of the order of the GUT scale, namely ${\cal W}\supset M_{GUT} (F^i_L\bar F^i_L+ F^i_R\bar F^i_R+ D^i_6\bar D^i_6)$.
For the Higgs fields given in~(\ref{H4b},\ref{H4}), however, there is an additional contribution due to
the coupling with the adjoint,
\be
{\cal W}_H \supset \ov{H}\Sigma H + M_H \ov{H}\, H~\cdot
\ee
Substituting (\ref{Ajvev}), we obtain
\be
{\cal W}_H\supset (\frac V3+M_H)\bar Q_H^c Q_H^c + (M_H-V) \bar L_H L_H~,
\label{Hsplit}
\ee
where $L^c_H=(e_H^c,\nu^c_H)^T, Q^c_H=(u_H^c,d^c_H)^T$ are $SU(2)_R$ doublets,
and $\bar L^c_H, \bar Q^c_H$ their complex conjugates.
In the two step breaking pattern  $L^c_H, \bar L^c_H$ must develop TeV vevs
 to break the  $SU(2)_R$ symmetry.  Therefore, choosing $V\simeq M_H\sim {\cal O}(M_{GUT})$,
  $L^c_H, L^c_H$ remain in the low energy spectrum while  $\bar Q_H^c Q_H^c $ acquire masses
$M_{Q_H}\sim \frac 43 M_{H}\sim{\cal O}(M_{GUT})$.
This   ``singlet-triplet'' splitting is similar to the ``doublet-triplet" splitting in $SU(5)$ scenario.

\section{Low Energy  Phenomenology }

In the previous sections we stressed that there is a natural way to
provide  masses for the vector-like states that  might appear in the  spectrum.
Hence, fields such as color sextets $D_6$ and triplets $\Delta_{L,R}$ of Table \ref{T1}
decouple from the light spectrum as long as they appear in pairs with opposite $U(1)$ charges.
We also noticed that, when the $SU(4)_C$ adjoint Higgs acquires its vev, the Higgs pair $H+\bar H$
is a remarkable exception to this rule and, therefore, it remains in the `massless' spectrum of the effective theory.

In this section we analyse the Yukawa sector and, in particular, the terms
generating masses for the SM charged fields and neutrinos. The superpotential
terms of the effective model  should be invariant under the gauge symmetry
of the original D-brane configuration of the model.  We will also
introduce `matter' parity ${\cal R}$ to suppress unwanted superpotential terms and possible exotic interactions.  To this end, the fields
$\Sigma, H, \bar H, \bar N, h$ acquire vevs and will be assigned positive ${\cal R}$ parity 
and the fields $F_L, \bar F_R, N$ are assigned negative  ${\cal R}$ parity.

Next, we analyse the contributions of the tree level terms. The third family charged fermion
and Dirac neutrino mass could  emerge, to a good approximation,  from a common Yukawa term, $\lambda\,\bar F_RF_Lh$.
 As a result, in the simplest constructions,
the 4-2-2 symmetry  imposes approximate third family Yukawa unification
$\lambda_t\simeq \lambda_b\simeq \lambda_{\tau}\simeq \lambda_{\nu_D}$.

However, these are not the only mass terms for the neutrino sector.
The following invariant couplings  also  involve the neutrino fields
\be
\lambda\bar F_R F_L h+\lambda_1 \bar F_R H N+\lambda_2 \frac{1}{M_*}{\bar N}^2 NN~,\label{sup}
\ee
where $M_*$ is the compactification scale, $M_* \gtrsim  M_{GUT}$ .
The term $\bar F_R H N$ suggests that the singlet field $N=(1,1,1)_{(0,0,-2)}$
can be interpreted as a sterile neutrino.
Therefore, assuming  a non-zero vev for the scalar component of the $\bar N$ singlet  (positively `charged' under $U(1)_R$),
eq. (\ref{sup}) yields
\ba
{\cal W}&=&\lambda_1\langle h_u\rangle \nu^c\nu + \lambda_2\langle H\rangle \nu^cN +\frac{\lambda_3}{M_*}{\langle \tilde{\bar N}\rangle}^2NN\nn\\
&=&m_D \nu^c\nu + V_R \nu^cN+\mu\,NN~,
\ea
where we have defined
\ba  m_D= \lambda_1\langle h_u\rangle,\; V_R=\lambda_2\langle H\rangle,\;
  \mu=\frac{\lambda_3}{M_*}{\langle \tilde{\bar N}\rangle}^2~\cdot\label{VEVs}
  \ea 
The new neutral state $N$ coupled to the ordinary neutrino fields with the  superpotential couplings~(\ref{sup}) gives rise to an inverse see-saw mechanism~\cite{Mohapatra:1986bd,Malinsky:2005bi}.
Assuming three $N$ singlet fields, in particular,  we obtain a $6\times 6$ matrix of the form
\ba
M_{\nu}&\sim&\left(
\begin{array}{lll}
 0 & m_D^T & 0\\
 m_D& 0 & V_R \\
 0 & V_R & \mu
\end{array}
\right)~\cdot \label{invseesaw}
\ea
The three   left handed neutrinos obtain eigenmasses of order
\be
m_{\nu}\approx \frac{m_D^2}{V_R^2}\mu~,\label{mnu}
\ee
which must be sufficiently light (of  order $10^{-1}-10^{-2}$ eV) to interpret the
 neutrino oscillations data. 

In order to provide an estimate, we  make the following
assumptions for the various scales involed in~(\ref{mnu}).
Observing that the singlet $\bar N$  carries $U(1)_R$ `charge', we first make the 
reasonable assumption that its vev is associated with the $SU(2)_R$ breaking scale
\[   \langle \bar N\rangle =\kappa V_R \sim {\cal O}(V_R)~,\]
where $\kappa $ is a dimensionless  constant. Then, from~(\ref{VEVs}) we find that the scale $\mu$ is 
 \[ \mu = \kappa^2 \frac{V_R^2}{M_*}~\cdot \]
 Substituting into (\ref{mnu}), we find
 \ba 
 m_{\nu}
 &\simeq& {\kappa^2 \frac{m_D^2}{M_*}}
 \ea
For $M_*$ of the order of the GUT scale and the Dirac mass $m_D$ 
of the electroweak mass scale, and with  $\kappa \gtrsim  1$, we can obtain the desired left-handed neutrino masses. 

The interesting fact here is that a new mass scale $\mu $ is effectively generated and it is associated with the ``sterile neutrino''  $N$.
Since  the light neutrino scale  computed from the inverse see-saw matrix does not depend on the details
of the $SU(2)_R$ breaking scale, we are free to choose $V_R$ (provided it satisfies the experimental bounds)
so that the emergent scale $\mu$  takes the desired value. Thus, a well motivated choice is related to the possible 
interpretation of the dark matter puzzle through the existence of a light (order  a few keV) sterile neutrino. \footnote{See 
for example	\cite{Dodelson:1993je,Merle:2013wta} and references therein. Also, our estimates are in agreement 
with recent constraints~\cite{Yunis:2018eux}
arising from from the Galactic center give a bound $m_{\nu_s}\le 15$ keV. }
 In general, the two eigenvalues of this matrix are almost degenerate. However, with the particular choice  of $SU(2)_R$ breaking
	scale	and  the existence of several neutral singlets, as is the case in  the present D-brane construction,
 a  light neutral state can emerge naturally.~\footnote{See \cite{Agashe:2018oyk}
 	and \cite{Brdar:2018sbk} for recent related work.}

As an example, assuming  that the $\mu$ parameter in (\ref{invseesaw}) represents a submatrix  of the extra singlets  
with a scale of the order 10 keV,  the scale $V_R$ is given by
\[ V_R\sim\sqrt{M_*/{\rm GeV}}\, 10^{-3}\,   {\rm GeV}   \gtrsim 10^6 {\rm GeV}~\cdot   \]
Working out the eigenvalues of the mass matrix (\ref{invseesaw}) one can see that it is plausible to 
obtain a light neutral state of mass ${\cal O}$(10 keV) which could be interpreted as a dark matter component
provided it is sufficiently long lived.

A major issue in many grand unified models is the problem of rapid  proton decay. 
The 4-2-2 symmetry does not contain the  gauge bosons that mediate dimension six baryon number violating processes. 
Then, the only source of proton decay is associated with dimension five or higher dimensional operators, 
related to graphs containing  SM color triplets and other states, and the predictions are very model dependent. 
Operators of this kind are of the form $QQQ\ell, QQu^c e^c, u^cd^cd^c\nu^c$ etc., and  may arise from non-renormalisable 
terms of the form $F_LF_LF_LF_L$, $F_LF_L\bar F_R\bar F_R$ and $\bar F_R\bar F_R\bar F_R\bar F_R$. Such couplings, however,
 are prevented from the additional abelian symmetries $U(1)_{L,R}$ in the present construction. 
Then, the only  available states that might mediate baryon violating diagrams, are the triplets descending from the sextets  which, in principle, receive masses of order $M_{GUT}$.  Non-renormalisable   Yukawa couplings  of the form $\frac{1}{M_{str}} F_LF_LD_6\nu_s$ and $\frac{1}{M_{str}} \bar F_R\bar F_R\bar D_6\,N$ give rise to  diagrams similar to those discussed in~\cite{Leontaris:2017vzz}, and their strength  is determined by the  vev of the scalar
component of the singlet $ \nu_s$.  
There are no constraints  on the possible values of $ \langle\tilde\nu_s\rangle$, but if this is taken to be close to the GUT scale, the proton  lifetime is estimated to be of order $10^{35}$-$10^{36}$ yr, which will be tested by the recently approved Hyper Kamiokande  experiment~\cite{Abe:2016ero}.

\section{Discussion and Summary}

In this letter we have analysed a class of 4-2-2 models within the framework of intersecting D-branes. The low energy spectrum contains additional particles that normally do not appear in the usual field theory based 4-2-2 models. 
We focused on breaking the 4-2-2 symmetry to the SM gauge group via an intermediate step  $SU(3)_C \times SU(2)_L\times SU(2)_R\times  U(1)_{B-L}$.  The first stage of the two step  breaking pattern is realised with the
$SU(4)$ adjoint vev which  does not transform under the $SU(2)_{L/R}$, and so these latter gauge symmetry factors 
are preserved. The second stage of symmetry breaking proceeds with the use of the $H+\bar H=(Q_H+L_H)+(\bar Q_H^c+\bar L_H^c)$ representations of the
4-2-2 group. Implementing a missing partner mechanism discussed in the text,  the right handed  doublets ${\bar L}_H^c+ L_H^c$  remain in the
low energy spectrum and develop TeV scale vevs
along their neutral directions  $\langle \nu_H^c\rangle$ and $\langle\bar \nu_H^c\rangle$ respectively, thereby breaking the $SU(2)_R$ symmetry.

The possibility of keV mass sterile neutrino as a potential new dark matter candidate is a striking example of this. The appearance of additional $U(1)$ symmetries which prevent rapid proton decay is certainly very helpful and a much desired feature for realistic model building.  Proton decay via higher dimensional operators can yield lifetime estimates on the order of $10^{35}$-$10^{36}$ yr which will be tested by the 
Hyper Kamiokande  experiment.

\vspace{.8cm}
{\bf Acknowledgements}. {G.K.L. would like to thank the Physics and Astronomy Department and Bartol Research
	Institute of the University of Delaware, 	for kind hospitality where part of this work has been done.
	Q.S. is supported in part by the DOE grant DE-SC0013880.}


\newpage


\begin{thebibliography}{99}


%\cite{Pati:1974yy}
\bibitem{Pati:1974yy}
J.~C.~Pati and A.~Salam,
%``Lepton Number as the Fourth Color,''
Phys.\ Rev.\ D {\bf 10} (1974) 275
Erratum: [Phys.\ Rev.\ D {\bf 11} (1975) 703].
%doi:10.1103/PhysRevD.10.275, 10.1103/PhysRevD.11.703.2
%%CITATION = doi:10.1103/PhysRevD.10.275, 10.1103/PhysRevD.11.703.2;%%
%4210 citations counted in INSPIRE as of 18 May 2017	

%\cite{Sirunyan:2017acf}
\bibitem{Sirunyan:2017acf}
A.~M.~Sirunyan {\it et al.} [CMS Collaboration],
%``Search for massive resonances decaying into $WW$, $WZ$, $ZZ$, $qW$, and $qZ$ with dijet final states at $\sqrt{s}=13\text{ }\text{ }\mathrm{TeV}$,''
Phys.\ Rev.\ D {\bf 97} (2018) no.7,  072006
%doi:10.1103/PhysRevD.97.072006
[arXiv:1708.05379].
%%CITATION = doi:10.1103/PhysRevD.97.072006;%%
%34 citations counted in INSPIRE as of 27 Nov 2018


	%\cite{ATLAS:2015nsi}
	\bibitem{ATLAS:2015nsi}
	G.~Aad {\it et al.} [ATLAS Collaboration],
	%``Search for new phenomena in dijet mass and angular distributions from $pp$ collisions at $\sqrt{s}=$ 13 TeV with the ATLAS detector,''
	Phys.\ Lett.\ B {\bf 754} (2016) 302
	%doi:10.1016/j.physletb.2016.01.032
	[arXiv:1512.01530].
	%%CITATION = doi:10.1016/j.physletb.2016.01.032;%%
	%132 citations counted in INSPIRE as of 29 Sep 2017



%\cite{Mohapatra:1974hk}
\bibitem{Mohapatra:1974hk}
  R.~N.~Mohapatra and J.~C.~Pati,
  ``Left-Right Gauge Symmetry and an Isoconjugate Model of CP Violation,''
  Phys.\ Rev.\ D {\bf 11} (1975) 566.
%  doi:10.1103/PhysRevD.11.566
  %%CITATION = doi:10.1103/PhysRevD.11.566;%%
  %1780 citations counted in INSPIRE as of 25 Mar 2016



  %\cite{Shafi:1977yb}
  \bibitem{Shafi:1977yb}
    Q.~Shafi and C.~Wetterich,
    ``Left-Right Symmetric Gauge Models and Possible Existence of a Neutral Gauge Boson with Mass in the PETRA-PEP Energy Range,''
    Phys.\ Lett.\ B {\bf 73} (1978) 65.
   % doi:10.1016/0370-2693(78)90173-9
    %%CITATION = doi:10.1016/0370-2693(78)90173-9;%%
    %53 citations counted in INSPIRE as of 25 Mar 2016

%\cite{Mohapatra:1979ia}
\bibitem{Mohapatra:1979ia}
R.~N.~Mohapatra and G.~Senjanovic,
``Neutrino Mass and Spontaneous Parity Violation,''
Phys.\ Rev.\ Lett.\  {\bf 44} (1980) 912.
%doi:10.1103/PhysRevLett.44.912
%%CITATION = doi:10.1103/PhysRevLett.44.912;%%
%4320 citations counted in INSPIRE as of 18 May 2017
%\cite{Senjanovic:2016pza}
\\
%\bibitem{Senjanovic:2016pza}
G.~Senjanovic and V.~Tello,
``Origin of Neutrino Mass,''
PoS PLANCK {\bf 2015} (2016) 141.
%doi:10.22323/1.258.0141
%%CITATION = doi:10.22323/1.258.0141;%%
%4 citations counted in INSPIRE as of 27 Nov 2018

%\cite{Lazarides:1980nt}
\bibitem{Lazarides:1980nt}
  G.~Lazarides, Q.~Shafi and C.~Wetterich,
  ``Proton Lifetime and Fermion Masses in an SO(10) Model,''
  Nucl.\ Phys.\ B {\bf 181}, 287 (1981).
%  doi:10.1016/0550-3213(81)90354-0
  %%CITATION = doi:10.1016/0550-3213(81)90354-0;%%
  %979 citations counted in INSPIRE as of 25 Mar 2016

%\cite{Pati:2017ysg}
\bibitem{Pati:2017ysg} 
J.~C.~Pati,
%``Advantages of Unity With SU(4)-Color: Reflections Through Neutrino Oscillations, Baryogenesis and Proton Decay,''
Int.\ J.\ Mod.\ Phys.\ A {\bf 32}, no. 09, 1741013 (2017)
%doi:10.1142/S0217751X17410135
[arXiv:1706.09531].
%%CITATION = doi:10.1142/S0217751X17410135;%%
%1 citations counted in INSPIRE as of 03 Aug 2017

%\cite{CMS:2017wsr}
\bibitem{CMS:2017wsr}
CMS Collaboration [CMS Collaboration],
%``Search for diboson resonances in the $2l2\nu$ final state,''
CMS-PAS-B2G-16-023.
%%CITATION = CMS-PAS-B2G-16-023;%%
\\
%\cite{Aad:2015owa}
%\bibitem{Aad:2015owa}
G.~Aad {\it et al.} [ATLAS Collaboration],
%``Search for high-mass diboson resonances with boson-tagged jets in proton-proton collisions at $ \sqrt{s}=8 $ TeV with the ATLAS detector,''
JHEP {\bf 1512} (2015) 055
%doi:10.1007/JHEP12(2015)055
[arXiv:1506.00962].
%%CITATION = doi:10.1007/JHEP12(2015)055;%%
%280 citations counted in INSPIRE as of 29 Sep 2017


%\cite{Antoniadis:1988cm}
\bibitem{Antoniadis:1988cm}
  I.~Antoniadis and G.~K.~Leontaris,
  ``A Supersymmetric $SU(4) \times O(4)$ Model,''
  Phys.\ Lett.\  B {\bf 216}, 333 (1989).
  %%CITATION = PHLTA,B216,333;%%


%\cite{Leontaris:2000hh}
\bibitem{Leontaris:2000hh}
  G.~K.~Leontaris and J.~Rizos,
  ``A Pati-Salam model from branes,''
  Phys.\ Lett.\ B {\bf 510} (2001) 295
  doi:10.1016/S0370-2693(01)00592-5
  [hep-ph/0012255].
  %%CITATION = doi:10.1016/S0370-2693(01)00592-5;%%
  %23 citations counted in INSPIRE as of 25 Mar 2016
  

 %\cite{Cvetic:2004ui}
 \bibitem{Cvetic:2004ui}
  M.~Cvetic, T.~Li and T.~Liu,
  ``Supersymmetric Pati-Salam models from intersecting D6-branes: A Road to the standard model,''
  Nucl.\ Phys.\ B {\bf 698} (2004) 163
 % doi:10.1016/j.nuclphysb.2004.07.036
  [hep-th/0403061].
  %%CITATION = doi:10.1016/j.nuclphysb.2004.07.036;%%
  %110 citations counted in INSPIRE as of 25 Mar 2016
  
  
  %\cite{Cvetic:2015txa}
  \bibitem{Cvetic:2015txa}
  M.~Cvetic, D.~Klevers, D.~K.~M.~Pe\~na, P.~K.~Oehlmann and J.~Reuter,
  %``Three-Family Particle Physics Models from Global F-theory Compactifications,''
  JHEP {\bf 1508} (2015) 087
%  doi:10.1007/JHEP08(2015)087
  [arXiv:1503.02068].
  %%CITATION = doi:10.1007/JHEP08(2015)087;%%
  %17 citations counted in INSPIRE as of 18 May 201

%\cite{Ibanez:2001nd}
\bibitem{Ibanez:2001nd}
  L.~E.~Ibanez, F.~Marchesano and R.~Rabadan,
  %``Getting just the standard model at intersecting branes,''
  JHEP {\bf 0111} (2001) 002
  doi:10.1088/1126-6708/2001/11/002
  [hep-th/0105155].
  %%CITATION = doi:10.1088/1126-6708/2001/11/002;%%
  %450 citations counted in INSPIRE as of 19 Feb 2017


%\cite{Dijkstra:2004cc}
\bibitem{Dijkstra:2004cc}
  T.~P.~T.~Dijkstra, L.~R.~Huiszoon and A.~N.~Schellekens,
  %``Supersymmetric standard model spectra from RCFT orientifolds,''
  Nucl.\ Phys.\ B {\bf 710} (2005) 3
 % doi:10.1016/j.nuclphysb.2004.12.032
  [hep-th/0411129].
  %%CITATION = doi:10.1016/j.nuclphysb.2004.12.032;%%
  %195 citations counted in INSPIRE as of 19 Feb 2017


 %\cite{Anastasopoulos:2010ca}
 \bibitem{Anastasopoulos:2010ca}
   P.~Anastasopoulos, G.~K.~Leontaris and N.~D.~Vlachos,
   ``Phenomenological Analysis of D-Brane Pati-Salam Vacua,''
   JHEP {\bf 1005} (2010) 011
  % doi:10.1007/JHEP05(2010)011
   [arXiv:1002.2937].
   %%CITATION = doi:10.1007/JHEP05(2010)011;%%
   %13 citations counted in INSPIRE as of 25 Mar 2016

  %\cite{Anastasopoulos:2006da}
  \bibitem{Anastasopoulos:2006da}
    P.~Anastasopoulos, T.~P.~T.~Dijkstra, E.~Kiritsis and A.~N.~Schellekens,
    %``Orientifolds, hypercharge embeddings and the Standard Model,''
    Nucl.\ Phys.\ B {\bf 759} (2006) 83
  %  doi:10.1016/j.nuclphysb.2006.10.013
    [hep-th/0605226].
    %%CITATION = doi:10.1016/j.nuclphysb.2006.10.013;%%
    %146 citations counted in INSPIRE as of 30 Jun 2016

%\cite{Kobayashi:2004ya}
\bibitem{Kobayashi:2004ya}
T.~Kobayashi, S.~Raby and R.~J.~Zhang,
%``Searching for realistic 4d string models with a Pati-Salam symmetry: Orbifold grand unified theories from heterotic string compactification on a Z(6) orbifold,''
Nucl.\ Phys.\ B {\bf 704} (2005) 3
%doi:10.1016/j.nuclphysb.2004.10.035
[hep-ph/0409098].
%%CITATION = doi:10.1016/j.nuclphysb.2004.10.035;%%
%230 citations counted in INSPIRE as of 22 Mar 2017

%\cite{Mohapatra:1986bd}cite{Malinsky:2005bi}
\bibitem{Mohapatra:1986bd}
R.~N.~Mohapatra and J.~W.~F.~Valle,
%``Neutrino Mass and Baryon Number Nonconservation in Superstring Models,''
Phys.\ Rev.\ D {\bf 34} (1986) 1642.
%doi:10.1103/PhysRevD.34.1642
%%CITATION = doi:10.1103/PhysRevD.34.1642;%%
%856 citations counted in INSPIRE as of 06 Oct 2017


%\cite{Malinsky:2005bi}
\bibitem{Malinsky:2005bi}
M.~Malinsky, J.~C.~Romao and J.~W.~F.~Valle,
%``Novel supersymmetric SO(10) seesaw mechanism,''
Phys.\ Rev.\ Lett.\  {\bf 95} (2005) 161801
%doi:10.1103/PhysRevLett.95.161801
[hep-ph/0506296].
%%CITATION = doi:10.1103/PhysRevLett.95.161801;%%
%196 citations counted in INSPIRE as of 03 Oct 2017



  
%\cite{urRehman:2006hu}
\bibitem{urRehman:2006hu}
M.~ur Rehman, V.~N.~Senoguz and Q.~Shafi,
%``Supersymmetric And Smooth Hybrid Inflation In The Light Of WMAP3,''
Phys.\ Rev.\ D {\bf 75} (2007) 043522
%doi:10.1103/PhysRevD.75.043522
[hep-ph/0612023].
%%CITATION = doi:10.1103/PhysRevD.75.043522;%%
%63 citations counted in INSPIRE as of 19 Mar 2017


    

 

%\cite{Anchordoqui:2016kmu}
\bibitem{Anchordoqui:2016kmu} 
L.~A.~Anchordoqui, I.~Antoniadis, H.~Goldberg, X.~Huang, D.~Lust and T.~R.~Taylor,
%``Minimal left-right symmetric intersecting D-brane model,''
Phys.\ Rev.\ D {\bf 95}, no. 2, 026011 (2017)
%doi:10.1103/PhysRevD.95.026011
[arXiv:1611.09785].
%%CITATION = doi:10.1103/PhysRevD.95.026011;%%



%\cite{Dodelson:1993je}
\bibitem{Dodelson:1993je}
S.~Dodelson and L.~M.~Widrow,
%``Sterile-neutrinos as dark matter,''
Phys.\ Rev.\ Lett.\  {\bf 72} (1994) 17
%doi:10.1103/PhysRevLett.72.17
[hep-ph/9303287].
%%CITATION = doi:10.1103/PhysRevLett.72.17;%%
%818 citations counted in INSPIRE as of 27 Sep 2018


%\cite{Merle:2013wta}
\bibitem{Merle:2013wta}
A.~Merle, V.~Niro and D.~Schmidt,
%``New Production Mechanism for keV Sterile Neutrino Dark Matter by Decays of Frozen-In Scalars,''
JCAP {\bf 1403} (2014) 028
%doi:10.1088/1475-7516/2014/03/028
[arXiv:1306.3996].
%%CITATION = doi:10.1088/1475-7516/2014/03/028;%%
%90 citations counted in INSPIRE as of 27 Sep 2018



%\cite{Yunis:2018eux}
\bibitem{Yunis:2018eux}
R.~Yunis, C.~R.~Arg\"uelles, N.~E.~Mavromatos, A.~Molin\'e, A.~Krut, J.~A.~Rueda and R.~Ruffini,
%``New constraints on sterile neutrino dark matter from the Galactic Center,''
arXiv:1810.05756 [astro-ph.GA].
%%CITATION = ARXIV:1810.05756;%%


%\cite{Agashe:2018oyk}
\bibitem{Agashe:2018oyk}
K.~Agashe, P.~Du, M.~Ekhterachian, C.~S.~Fong, S.~Hong and L.~Vecchi,
%``Hybrid seesaw leptogenesis and TeV singlets,''
Phys.\ Lett.\ B {\bf 785} (2018) 489
%doi:10.1016/j.physletb.2018.09.006
[arXiv:1804.06847 [hep-ph]].
%%CITATION = doi:10.1016/j.physletb.2018.09.006;%%
%2 citations counted in INSPIRE as of 27 Sep 2018


%\cite{Brdar:2018sbk}
\bibitem{Brdar:2018sbk} 
V.~Brdar and A.~Y.~Smirnov,
%``Low Scale Left-Right Symmetry and Naturally Small Neutrino Mass,''
arXiv:1809.09115 [hep-ph].
%%CITATION = ARXIV:1809.09115;%%



%\cite{Babu:1998wi}
\bibitem{Babu:1998wi}
K.~S.~Babu, J.~C.~Pati and F.~Wilczek,
%``Fermion masses, neutrino oscillations, and proton decay in the light of Super-Kamiokande,''
Nucl.\ Phys.\ B {\bf 566} (2000) 33
%doi:10.1016/S0550-3213(99)00589-1
[hep-ph/9812538].
%%CITATION = doi:10.1016/S0550-3213(99)00589-1;%%
%314 citations counted in INSPIRE as of 16 Oct 2018

%\cite{Leontaris:2017vzz}
\bibitem{Leontaris:2017vzz}
G.~K.~Leontaris and Q.~Shafi,
%``Phenomenology with F-theory SU(5),''
Phys.\ Rev.\ D {\bf 96} (2017) no.6,  066023
%doi:10.1103/PhysRevD.96.066023
[arXiv:1706.08372].
%%CITATION = doi:10.1103/PhysRevD.96.066023;%%
%1 citations counted in INSPIRE as of 17 Oct 2018


%\cite{Abe:2016ero}
\bibitem{Abe:2016ero}
K.~Abe {\it et al.} [Hyper-Kamiokande Collaboration],
%``Physics potentials with the second Hyper-Kamiokande detector in Korea,''
PTEP {\bf 2018} (2018) no.6,  063C01
%doi:10.1093/ptep/pty044
[arXiv:1611.06118].
%%CITATION = doi:10.1093/ptep/pty044;%%
%70 citations counted in INSPIRE as of 16 Oct 2018

\end{thebibliography}
\end{document}